\documentclass{elsart}
\usepackage{latexsym}
\usepackage{alltt}
\usepackage[all]{xy}
\CompileMatrices
\xyoption{frame}
\usepackage{amssymb}
\usepackage{amscd}
\usepackage{amsmath}
\usepackage{theorem}
\theoremstyle{plain}
\newtheorem{theorem}{Theorem}[section]
\newtheorem{corollary}[theorem]{Corollary}
\newtheorem{lemma}[theorem]{Lemma}
\newtheorem{proposition}[theorem]{Proposition}
\newtheorem{definition}[theorem]{Definition}
{\theorembodyfont{\rmfamily} }
{\theorembodyfont{\rmfamily} \newtheorem{examples}[theorem]{Examples}}

\begin{document}
\begin{frontmatter}
\title{The structure of finite meadows}
\author[label1]{Inge Bethke\thanksref{email1}\corauthref{cor}}
\author[label1]{Piet Rodenburg\thanksref{email2}}
\author[label1]{Arjen Sevenster\thanksref{email3}}
\corauth[cor]{Corresponding author. Address: Kruislaan 403, 1098 SJ Amsterdam, The Netherlands}
\thanks[email1]{E-mail:  \texttt{I.Bethke@uva.nl}}
\thanks[email2]{E-mail:  \texttt{P.H.Rodenburg@uva.nl}}
\thanks[email3]{E-mail:  \texttt{A.Sevenster@uva.nl}}
\address[label1]{University of Amsterdam, Faculty of Science, Section Theoretical
Software Engineering (former Programming Research Group)}
\begin{abstract}
A \emph{meadow} is a commutative ring with a total inverse operator satisfying $0^{-1}=0$. We show that the
class of finite meadows
is the closure of the class of Galois fields under finite products. As a corollary, we obtain a unique
representation of minimal finite meadows in terms of finite prime fields.
\end{abstract}


\end{frontmatter}

\section{Introduction}
In abstract algebra, a field is a structure with operations of addition, subtraction
and multiplication. Moreover, every element has a multiplicative inverse|except 0. In a field, the rules
hold
which are familiar from the arithmetic of ordinary numbers.
The prototypical example is the field of rational numbers. Fields  can be specified by the axioms for
commutative rings with identity element, and the negative conditional formula
\[
x\neq 0 \rightarrow x\cdot x^{-1}=1,
\]
which is difficult to apply or automate in formal reasoning.

The theory of fields is a very active area which is not only of great theoretical interest but has also found
applications both within mathematics|combina\-to\-rics and algorithm analysis|as well as in engineering sciences
and, in particular, in coding theory and
sequence design. Unfortunately, since fields are not
axiomatized by equations only, \emph{Birkhoff's Theorem} fails, i.e. fields do not
constitute a variety: they are not closed under products, subalgebras, and homomorphic images.
In \cite{BT07}, the concept of \emph{meadows} was introduced, structures very similar to fields|the
considerable difference being that meadows do form a variety.

All fields and products of fields can be viewed as meadows|basically by stipulating $0^{-1}=0$|but not conversely.
 Also, every commutative \emph{Von Neumann regular ring} (see e.g.\ \cite{G79}) can be expanded
to a meadow (cf.\ \cite{BHT07}).

The aim of this paper is to describe the structure of finite meadows.
We will show that the class of finite meadows is the closure
of the class of finite fields under finite products. As a corollary, we obtain a unique representation of
minimal meadows in terms of prime fields. This result also follows from the observation that meadows are
\emph{biregular} and hence \emph{semisimple} rings, and the connection between commuting idempotents and direct product
decomposition into simple rings as   expounded in \cite{DH66}. Here, however, we will give a direct proof by a straightforward combination
of basic properties of meadows.

\section{Preliminaries}
In this section we recall the basic properties of rings and meadows.
\begin{definition}
A \emph{commutative ring}  is a structure $\langle R,+,-,\cdot, 0,1 \rangle$ such that
for all $x,y,z \in R$
\[
\begin{array}{lrcl}
(1)&(x+y)+z &=&x+(y+z)\\
(2)&x+y&=&y+x\\
(3)&x +0 &=& x\\
(4)&x+(-x)&=&0\\
(5)&(x \cdot y)\cdot z&=& x\cdot (y \cdot z)\\
(6)&x\cdot y &=&y \cdot x\\
(7)&x \cdot 1 &=& x\\
(8)&x\cdot (y + z) &=& x\cdot y + x \cdot z.
\end{array}
\]
We will write $x-y$ for $x+(-y)$.
\end{definition}
The following properties of commutative rings are well-known.
\begin{proposition}\label{basicprops}
Let $R$ be a commutative ring and $x,y \in R$. Then
\begin{enumerate}
\item the identity 1 is unique,
\item $0\cdot x=0$,
\item $(-x)\cdot y=-(x\cdot y)$,
\item $(-1)\cdot x = -x$
\item $- 0 = 0$,
\item $(-x) + (-y) = -(x+y)$,
\item $-(-x)=x$.
\end{enumerate}
\end{proposition}
\begin{proposition}\label{unique}
Let $R$ be a commutative ring.  For any $x\in R$, there exists at most one $y\in R$ with $x\cdot x \cdot y=x$ and
$y\cdot y \cdot x = y$.
\end{proposition}
{\bf Proof:}
Let $z$ be another element such that $x\cdot x\cdot z= x$ and $z\cdot z \cdot x= z$.
We have 
\[
y=y\cdot y\cdot x=y\cdot y\cdot (x\cdot x\cdot z) = (y\cdot y \cdot x)\cdot (x \cdot z)= 
y \cdot x \cdot z = x \cdot y \cdot z.
\]
Hence, by symmetry, $z=x\cdot y\cdot z$ and thus $y=z$.
 \hfill $\qed$
\begin{definition}
Let $R$ be a commutative ring and $x\in R$. If it exists, we call the element $y\in R$ uniquely
determined by $x \cdot x \cdot y = x $ and $y \cdot y \cdot x = y$ the
\emph{generalized inverse of $x$} and denote it by $x^{-1}$.
\end{definition}
\begin{proposition}
Ler $R$ be a commutative ring. We have
\begin{enumerate}
\item $0^{-1}=0$
\item $1^{-1}=1$ and $(-1)^{-1} = -1$
\item $(x^{-1})^{-1}= x$ for all $x\in R$ for which the generalized inverse exists.
\end{enumerate}
\end{proposition}
{\bf Proof:} (1) From $0\cdot 0\cdot 0 = 0$ it follows that  $0$
is the generalized inverse of 0, i.e.\  $0^{-1}=0$. (2) From
$1\cdot 1\cdot 1 = 1$ it follows that  $1$ is the generalized
inverse of 1, i.e.\  $1^{-1}=1$, and similarly $(-1)^{-1}=-1$. (3)
Since the equalities $x\cdot x \cdot a =x$ and $a\cdot a \cdot x
=a$ are symmetric in $a$ and $x$, it follows that $x$ is the
inverse of $a$. Thus $x = a^{-1}= (x^{-1})^{-1}$. \hfill $\qed$

\begin{examples}\label{example}
\begin{enumerate}
\item In the commutative ring $\mathbb{Q}$ of rational numbers, every element has a
generalized inverse. If $x\neq 0$, the inverse is just  the ``regular" inverse,
and $0^{-1}=0$.
\item
Consider the ring $\mathbb{Z}/10\mathbb{Z}$ with elements $\{0,1,2, \ldots , 9\}$ where arithmetic is performed modulo $10$.
We find that every element has a generalized inverse as follows:
\[
\begin{array}{rclccrcl}
(0)^{-1} &=& 0&\hspace{1 cm}&
(1)^{-1} &=& 1\\
(2)^{-1} &=& 8&&
(3)^{-1} &=& 7\\
(4)^{-1} &=& 4&&
(5)^{-1} &=& 5\\
(6)^{-1} &=& 6&&
(7)^{-1} &=& 3\\
(8)^{-1} &=& 2&&
(9)^{-1} &=& 9\\
\end{array}
\]
Note that the equation $2\cdot 2\cdot x =2\ mod\ 10$ has two
solutions: the generalized inverse of 2, namely 8, and the
``pseudo" inverse 3 which does not satisfy the equation $x \cdot x
\cdot 2 =x \ mod \ 10$.
\item Consider $\mathbb{Z}/4\mathbb{Z}$. We find that $0$, $1$ and $3$ have generalized
inverses, namely $0^{-1}=0$,  $1^{-1}=1 $, and $3^{-1}=3$, but
that 2 has no generalized inverse, because the equation $2\cdot
2\cdot x = 2 \ mod \ 4$ has no solutions.
\end{enumerate}
\end{examples}
\begin{definition}
A \emph{meadow} is a commutative ring in which every element has a generalized inverse. 
\end{definition}
\begin{examples}
$\mathbb{Q}$ and $\mathbb{Z}/10\mathbb{Z}$ are meadows, $\mathbb{Z}/4\mathbb{Z}$
is not a meadow.
\end{examples}
\begin{proposition}
Let $M$ be a meadow. For $x,y \in M$ we have
\begin{enumerate}
\item $x \cdot x^{-1} = 0 \Leftrightarrow x=0$
\item $x \cdot y = 1 \Rightarrow x^{-1}= y$
\item $(x\cdot y)^{-1} = x^{-1}\cdot y^{-1}$
\item $(-x)^{-1}= -(x^{-1})$
\end{enumerate}
\end{proposition}
{\bf Proof:} (1) If $x\cdot x^{-1} = 0$, then $x=x\cdot x \cdot x^{-1}= x\cdot 0= 0$.
The converse follows from Proposition \ref{basicprops}.(2). (2) If $x \cdot y =1$, then
$x \cdot x \cdot y = x\cdot 1 = x$ and $y \cdot y \cdot x= y\cdot 1= y$. Hence
$y = x^{-1}$, since the generalized inverse is uniquely determined. (3) We have to show that  the generalized inverse of $x\cdot y$
equals $x^{-1}\cdot y^{-1}$. We have
\[
(x\cdot y)\cdot (x\cdot y) \cdot (x^{-1}\cdot y^{-1}) = (x\cdot x \cdot x^{-1}) \cdot
(y\cdot y \cdot y^{-1})= x \cdot y
\]
and
\[
(x^{-1}\cdot y^{-1})\cdot (x^{-1}\cdot y^{-1}) \cdot (x\cdot y) =
(x^{-1}\cdot x^{-1} \cdot x) \cdot
(y^{-1}\cdot y^{-1} \cdot y) =
\]
\[
(x^{-1}\cdot x^{-1} \cdot (x^{-1})^{-1}) \cdot
(y^{-1}\cdot y^{-1} \cdot (y^{-1})^{-1}) = x^{-1}\cdot y^{-1}
\]
and the result follows from unicity of the  generalized inverse. (4)
$(-x)^{-1}=(-1\cdot x)^{-1}=(-1)^{-1} \cdot x^{-1}=-1\cdot x^{-1}=-x^{-1}$.
\hfill $\qed$
\begin{proposition}\label{power}
Let $M$ be a meadow. For $x\in M$ we have
\begin{enumerate}
\item $x^2=x \Rightarrow x^{-1}=x$
\item for $n>2$, $x^n=x \Rightarrow x^{-1}=x^{n-2}$.
\end{enumerate}
\end{proposition}
{\bf Proof:} (1) To prove that $x$ is its own inverse, it suffices
to prove $x\cdot x\cdot x = x$. We have $x\cdot x\cdot x = x\cdot
x =x$. (2) If $n=3$ we have $x^3=x$, hence $x^{-1}=x$.
If $n>3$, we have $x\cdot x \cdot x^{n-2} = x$ and
\[
x^{n-2}\cdot x^{n-2}\cdot x=x^{2n-4}\cdot x=x^n\cdot x^{n-4}\cdot x=x\cdot x^{n-4}\cdot x=x^{n-2}.
\]
Hence $x^{-1}= x^{n-2}$. \hfill $\qed$
\begin{proposition}
$\mathbb{Z}/n\mathbb{Z}$ is a meadow if and only if $n$ is
squarefree, i.e. $n$ is the product of pairwise distinct primes.
\end{proposition}
{\bf Proof:} Let $\mathbb{Z}/n\mathbb{Z}$ be a meadow. Then the
equations
\[
a^2x\equiv a \ mod\ n\ \text{ and }x^2 a\equiv x\ mod\ n\
\]
have a unique solution for all $a$. Suppose $p^{\alpha} \mid n$
with $p$ prime and $\alpha \geq 1$. Taking $a=p$ in the first
equation, we
conclude that $\alpha = 1$.\\
Conversely, let $n$ be squarefree. Note that this implies
$(a^2,n)=(a,n)$ for all $a$. First assume $(a,n)=1$. Then we conclude
from $(a^2,n)=(a,n)=1$ that $a^2x\equiv a\ mod\ n$ has a  unique solution,
say $\xi$, i.e.\ $n\mid a^2\xi -a=a(a\xi-1)$ and therefore $n\mid
a\xi -1$ since $(a,n)=1$. Hence $n\mid \xi(a\xi -1)=a\xi^2-\xi$,
i.e.\ $\xi$ is a solution of $x^2a\equiv x\ mod\ n$ as well. 
Now let $(a,n)>1$.  To minimize notation let us assume that $n=pq$ with
$p$ and $q$ different primes.  Then $(a,n)=p$ or $(a,n)=q$. So let us assume
$(a,n) =p$ and put $a =\alpha p$, where obviously $q\nmid \alpha$.  From $a^2x \equiv a\ mod\ n$
we get $\alpha^2p^2x \equiv \alpha p\ mod\ pq$, i.e.\ $\alpha^2px \equiv \alpha \ mod\ q$.
Since $(\alpha^2p,q)=1$ this equation has exactly one solution $\xi$ and the $p$ solutions of
$a^2x\equiv a\ mod\ n$ are represented by $\xi, \xi + q, \ldots , \xi + (p-1)q$. Let $\xi ' $ be the solution divisible
 by $p$. Then it is easy to check that $\xi '$ is also a solution of $x^2a\equiv x \ mod\ n$.
\hfill $\qed$

Let us note that this proposition also follows directly from our main result Corollary \ref{decomp}.

\section{Decomposition of finite meadows}
In \cite{B44} it is proved that every commutative regular ring in the sense of von Neumann
is a subdirect union of fields.
In this section we show that every finite meadow is a direct product of
finite fields. Part of the proof is also known from the theory of rings:
under certain conditions|also met in our case|a ring $R$ can be decomposed as
$R=e_1\cdot R\cdot e_1 \oplus \ldots \oplus e_n\cdot R\cdot e_n$ where $\{e_1, \ldots ,
e_n\}$ is the set of mutually orthogonal minimal idempotents in $R$ (see e.g.\ \cite{D06}).
\begin{definition}
Let $M$ be a meadow.
\begin{enumerate}
\item An element $e\neq 0$ in $M$ is an \emph{idempotent} if $e \cdot e = e$.
\item If $e, e'\in M$ are idempotents then we write $e \leq e'$ if $e\cdot e'=e$.
\item An idempotent $e\in M$ is \emph{minimal} if for every idempotent
$e'\in M$,
\[e'\leq e \Rightarrow e'=e.\]
\end{enumerate}
\end{definition}
\begin{proposition}\label{meadow}
Let $M$ be a meadow and $e\in M$ an idempotent.
Then
\begin{enumerate}
\item $e=e^{-1}$
\item $e\cdot M$ is a meadow with multiplicative identity element $e$.
\item If $e$ is minimal then $e\cdot M$ is a field with multiplicative identity element $e$.
\end{enumerate}
\end{proposition}
{\bf Proof:}
\begin{enumerate}
\item This is Proposition \ref{power} (1).
\item Since idempotents are self-inverse $e\cdot M$ is closed under $+, \cdot, \-^{-1}$ and clearly satisfies the
 axioms for meadows.
\item Since $e\cdot M$ is a meadow with multiplicative identity element $e$, it suffices to prove that
$(e\cdot m)\cdot (e\cdot m)^{-1} = e$ for every $e\cdot m \neq 0$. Thus let $e\cdot m$ be a nonzero element.
Then $(e\cdot m)\cdot (e\cdot m)^{-1} \neq 0$ because otherwise
 \[
e\cdot m = (e\cdot
 m)\cdot (e \cdot m) \cdot (e\cdot m)^{-1} = 0.
 \]
 Moreover,
\[
(e\cdot m)\cdot (e\cdot m)^{-1}\cdot (e\cdot m)\cdot (e\cdot m)^{-1} = (e\cdot m)\cdot (e\cdot m)^{-1}.
\]
So $(e\cdot m)\cdot (e\cdot m)^{-1}$ is an idempotent. Hence, since
\[
e\cdot (e\cdot m)\cdot (e\cdot m)^{-1} = (e\cdot m)\cdot (e\cdot m)^{-1}
\]
and $e$ is minimal we have $(e\cdot m)\cdot (e\cdot m)^{-1} = e$.
\end{enumerate}
\hfill $\qed$

The main properties of idempotents are summarized in the following proposition.
\begin{proposition}\label{properties}
Let $M$ be a meadow.
\begin{enumerate}
\item $\leq$ is a partial order on the idempotents.
\item If $e,e'\in M$ are idempotents and $e\cdot e'\neq 0$ then $e\cdot e'$ is also an idempotent.
\item If $e,e'\in M$ are  idempotents and $e < e'$ then $e'-e$ is also an idempotent.
\end{enumerate}
\end{proposition}
{\bf Proof}:
\begin{enumerate}
\item Clearly $\leq$ is reflexive. If $e\leq e'$ and $e'\leq e''$ then
\[
e\cdot e'' = (e\cdot e') \cdot e'' = e\cdot (e' \cdot e'') = e\cdot e' =e.
\]
Therefore the relation is transitive. Finally, if $e\leq e'$ and $e'\leq e$ then
\[
e = e\cdot e' = e'\cdot e=e.'
\]
Thus $\leq$ is also antisymmetric.
\item We multiply $e\cdot e'$ with itself:
$
(e\cdot e')\cdot (e\cdot e') = (e\cdot e)\cdot (e'\cdot e') =e\cdot e'.
$
\item We multiply $e'-e$ with itself:
\[
(e'-e)\cdot (e'-e)= e'\cdot e' -e\cdot e' -e'\cdot e +e\cdot e = e' -e -e +e = e'-e.
\]
\end{enumerate}
\hfill $\qed$
\begin{definition}
Let $M$ be a meadow and
$e,e' \in M$. We call $e$ and $e'$ \emph{orthogonal} if $e\cdot e'=0$.
\end{definition}
\begin{proposition}\label{orthogonal}
Let $M$ be a meadow.
\begin{enumerate}
\item If $e,e'\in M$ are different minimal idempotents then $e$ and $e'$ are orthogonal.
\item If $e,e'\in M$ are orthogonal idempotents then $e+e'$ is an idempotent.
\end{enumerate}
\end{proposition}
{\bf Proof}:
\begin{enumerate}
\item Suppose $e\cdot e'\neq 0$. Then $e\cdot e'$ is an idempotent by Proposition \ref{properties}(2). Moreover,
$
e\cdot e' = e\cdot e \cdot e' = e\cdot e' \cdot e,
$
i.e.\ $e\cdot e' \leq    e$. Thus $e\cdot e' = e$, since $e$ is minimal. Likewise $e\cdot e' = e'$ and hence
$e=e'$. Contradiction.
\item We multiply again:
\[
(e + e')\cdot (e+ e') = e\cdot e + e \cdot e' + e' \cdot e + e' \cdot e' = e + 0
+ 0 + e' =e + e'.
\]
Moreover, $(e + e') \cdot e = e\cdot e + e\cdot e'= e$. Hence $e+e'\neq 0$.
\end{enumerate}
\hfill $\qed$

We now show that every finite meadow is the direct product of the fields generated by its minimal
idempotents.
\begin{lemma}\label{som}
Let $M$ be a finite meadow and $\{e_1, \ldots ,e_n\} \subseteq M$ be the set of minimal
idempotents. Then $e_1 + \cdots  + e_n = 1$.
\end{lemma}
{\bf Proof}: Since minimal idempotents are orthogonal we have $e_i\cdot e_j = 0$
for $i\neq j$ by Proposition \ref{orthogonal} (1). Therefore for every $1 \leq i < n$, $e_1 + \cdots + e_i$ is an
idempotent orthogonal with $e_{i+1}$, and hence $e_1 + \cdots  + e_n$ is an idempotent by
Proposition \ref{orthogonal} (2). And therefore $1 - (e_1 + \cdots  + e_n)$ is an idempotent by Proposition
\ref{properties} (3) unless it is 0. Suppose $1 - (e_1 + \cdots  + e_n)$ is an idempotent. Then, since $\leq$ is
a partial order there must be some minimal idempotent $e_i\leq 1 - (e_1 + \cdots  + e_n)$. But
\[
\begin{array}{rcl}
e_i \cdot (1 - (e_1 + \cdots  + e_n))& =& e_i -(e_i\cdot e_1 + \cdots + e_i\cdot e_i + \cdots + e_i\cdot e_n)\\
&= &e_i - (0 + \cdots + e_i\cdot e_i + \cdots + 0)\\
& = &0
\end{array}
\]
Contradiction. Hence $1 - (e_1 + \cdots  + e_n)$ is not an idempotent, i.e.
\[
1 - (e_1 + \cdots  + e_n)=0
\]
whence $e_1 + \cdots  + e_n = 1$. \hfill $\qed$

\begin{theorem}
Let $M$ be a finite meadow and $\{e_1, \ldots ,e_n\} \subseteq M$ the set of minimal
idempotents. Then
\[
M \cong e_1\cdot M \times \cdots \times e_n \cdot M
\]
\end{theorem}
{\bf Proof}: Because the theory of meadows is equational, we know from universal algebra that
a direct product of meadows is a meadow.
Thus $e_1\cdot M \times \cdots \times e_n \cdot M$ is a meadow with multiplicative
identity element
$(e_1, \ldots , e_n)$ and the operations defined componentwise.
Define $h: M \rightarrow e_1\cdot M \times \cdots \times e_n \cdot M$ by
\[
h(m) = (e_1\cdot m, \ldots , e_n\cdot m).
\]
 Then $h$ is a homomorphism.
Suppose $h(m)=h(m')$. Then for every $1\leq i\leq n$, $e_i\cdot m =e_i\cdot m'$.
Thus
\[
\begin{array}{rcl}
m=1\cdot m&=& (e_1 + \cdots +e_n)\cdot m\\
&=& e_1\cdot m +\cdots +e_n \cdot m \\
&=&e_1\cdot m' +\cdots +e_n \cdot m' \\
&= &(e_1 + \cdots +e_n)\cdot m' = 1\cdot m'=m'.
\end{array}
\]
Hence $h$ is injective. Now let $(e_1\cdot m_1, \ldots , e_n\cdot
m_n)\in e_1\cdot M \times \cdots \times e_n \cdot M$ and consider
$m= e_1\cdot m_1 + \ldots + e_n\cdot m_n$. Then, since $e_i$ and
$e_j$ are orthogonal for $i\neq j$, $e_i\cdot m = e_i \cdot m_i$.
Thus $h(m)=(e_1\cdot m_1, \ldots , e_n\cdot m_n)$. Whence $h$ is
also surjective. \hfill $\qed$

The order, or number of elements, of finite fields
is of the form $p^n$, where $p$ is a prime number.  Since any two finite
fields with the same number of elements are isomorphic, there is a naming
scheme of finite fields that specifies only the order of the field.
One notation for a finite field|or more precisely, its zero-totalized
expansion, in which inverse is a total operation with $0^{-1}=0$|with $p^n$ elements is $GF(p^n)$,
where the letters $GF$ stand for \emph{Galois field}. From the above theorem
it now follows immediately that the class of finite meadows is
the closure of  the class of Galois fields under finite products.
\begin{corollary}\label{decomp}
Let $M$ have $n$ elements. Then $M$ is a meadow if and only if
there are|not necessarily distinct|primes $p_1, \ldots , p_k$ and
natural numbers $n_1, \ldots , n_k$ such that
\[
M \cong GF(p_1^{n_1})\times \cdots \times GF(p_k^{n_k})
\]
and $n= p_1^{n_1} \cdots p_k^{n_k}$.
\end{corollary}
Observe that|as a consequence|meadows of the same size are not necessarily isomorphic:
$GF(4)$ and $GF(2)\times GF(2)$ are both meadows but $GF(4)\not \cong GF(2)\times GF(2)$.
\begin{definition}
A meadow is \emph{minimal} if it does not contain a proper submeadow.
\end{definition} 
\begin{corollary}
\begin{enumerate}
\item Let $M$ be a finite  meadow with cardinality $n$. Then
$M$ is minimal if and only if there exist distinct primes $p_1, \ldots
, p_k$ such that  
\[
M \cong \mathbb{Z}/p_1 \mathbb{Z} \times \cdots \times \mathbb{Z}/p_k \mathbb{Z}
\]
and $n= p_1 \cdots p_k$.
\item Finite minimal meadows of the same size are isomorphic.
\end{enumerate}
\end{corollary}
{\bf Proof}: (2) follows from (1) and (1) follows from the
preceding corollary.  \hfill $\qed$

As an application of Corollary \ref{decomp}, we determine the number of self-inverse
and invertible elements in finite meadows.
\begin{definition}
Let $M$ be a meadow and $m\in M$. Then
\begin{enumerate}
\item $m$ is \emph{self-inverse} if $m = m^{-1}$,
\item $m$ is \emph{invertible} if $m \cdot m^{-1} = 1$,
\end{enumerate}
\end{definition}
So, e.g.\ in $\mathbb{Z}/10\mathbb{Z}$ (see Example
\ref{example}(2)) $0,1,4,5,6$ are self-inverse elements, $1, 3,7,$
are invertibles, and $9$ is both self-inverse and invertible.
\begin{proposition}
Let $M\cong GF(p_1^{k_1})\times \cdots \times GF(p_n^{k_n})$. Then
$M$ has
\begin{enumerate}
\item $2^l\cdot 3^{n-l}$ self-inverses, where $l=\mid\{i\mid 1\leq i\leq n \ \&\
p_i = 2\ \& \ k_i=1\}\mid$, and
\item $(p_1^{k_1}-1) \cdots  (p_n^{k_n}-1)$ invertibles.
\end{enumerate}
\end{proposition}
{\bf Proof}: First observe that the number of self-inverses [invertibles] of $M$ is the product of
the number of self-inverses [invertibles] in the Galois fields.\\
(1) Now $m$ is self-inverse in a meadow iff $m^3=m\cdot m\cdot m^{-1}=m$. Thus the number of self-inverses in
$GF(p_i^{k_i})$ is the number of elements such that $m\cdot (m-1)\cdot (m+1) = 0$.
Since a field has no zero divisors, these are precisely the elements $0,1$ and $-1$.
Thus if $p_i=2$ and $k_i=1$ then $GF(p_i^{k_i})$ has 2 self-inverses and otherwise 3.\\
(2) Since in a field every element is invertible except $0$,
$GF(p_i^{k_i})$ has $p_i^{k_i}- 1$ invertibles. \hfill $\qed$
\section{Skew meadows}
Skew meadows differ from meadows only in that their multiplication is not required to be commutative. We 
here deviate from the exposition given in \cite{BHT09} and give a
slightly
different but equivalent definition.
\begin{proposition}
Let $R$ represent a noncommutative ring with identity 1, i.e. such that $1\cdot x = x \cdot 1 = x$ for every $x\in R$. If 
for $x\in R$ there exists
a $y\in R$ such that
\begin{enumerate}
\item $x\cdot x\cdot y =x$, 
\item $y \cdot y\cdot x = y$,
\item $x \cdot y\cdot y = y$, and
\item $y \cdot x \cdot x = x$
\end{enumerate}
then $y$ is unique.
\end{proposition}
{\bf Proof}: As in Proposition \ref{unique}. \hfill $\qed$.
\begin{definition}
Let $R$ be a noncommutative ring with identity 1.
\begin{enumerate}
\item Let $x\in R$. If it exists, we call the element $y\in R$ uniquely determined by the equations (1)--(4) in the previous 
definition the 
\emph{generalized inverse of} $x$
and denote it by $x^{-1}$.
\item If every element in $R$ has a generalized inverse, then $R$ is called a \emph{skew meadow}.
\end{enumerate}
\end{definition}
By the proof of \cite{BHT09} (Theorem 4.13), every skew meadow is a subdirect product of zero-totalized devision rings. Hence a finite
skew meadow is a subdirect product of zero-totalized finite devision rings; by Wedderburn's Theorem, these are fields. So
every finite skew meadow is commutative.

\section{Conclusion}
We have described the finite meadows as follows:
\begin{enumerate}
\item the class of finite meadows is the closure of the class of Galois fields under finite products,
\item in contrast with finite fields, finite meadows of the same size are not necessarily isomorphic; however,
\item minimal finite meadows of the same size are unique up to isomorphism.
\end{enumerate}
This gives a clear picture of the finite objects in the category of meadows.

\mbox{}\\
\mbox{}\\
{\bf Acknowledgement}: We are indebted to one of the referees of an earlier version of this paper
who observed that our results on invertibles and self-inverses are simple corollaries of the Chinese Remainder Theorem
and ring decomposition. \\
This work was partly supported by The Netherlands Organisation for Scientific Research (NWO) under grant 638.003.611.

\end{document}